\documentclass[aps,prd,onecolumn,groupedaddress,showpacs]{revtex4}

\usepackage{amsmath}
\usepackage{graphicx}

\begin{document}

\title{What can we learn from a measurement of $\sin(2 \beta + \gamma)$?}

\author{Jo\~{a}o P.\ Silva}
\affiliation{Centro de F\'{\i}sica das Interac\c{c}\~{o}es Fundamentais,
	Instituto Superior T\'{e}cnico,
	P-1049-001 Lisboa, Portugal,}
\affiliation{Instituto Superior de Engenharia de Lisboa,
	Rua Conselheiro Em\'{\i}dio Navarro,
	1900 Lisboa, Portugal.}

\author{Abner Soffer}
\affiliation{Department of Physics, 
	Colorado State University, Fort Collins, CO 80523, USA.}

\author{Lincoln Wolfenstein}
\author{Feng Wu}
\affiliation{Carnegie Mellon University, Department of Physics,
	5000 Forbes Avenue, Pittsburgh, PA 15213, USA.}

\date{\today}

\begin{abstract}
The constraints on the value of the CKM phase $\gamma$
that may be achieved by prospective measurements of
$\sin{2 \beta}$ and $\sin{(2 \beta + \gamma)}$ are discussed.
Significant constraints require quite small errors,
and may depend on assumptions about strong phases.
The measurement of $\sin{(2 \beta + \gamma)}$ combined with
other experiments could provide valuable limits on new physics
in $B_d - \overline{B_d}$ mixing.
\end{abstract}

\pacs{11.30.Er, 13.25.Ft, 13.25.Hw, 14.40.-n.}

\maketitle

\section{\label{sec:intro}Introduction}

The major goal of $B$ physics is to provide quantitative tests
of the Standard Model description of the charged current interactions,
through the Cabibbo--Kobayashi--Maskawa (CKM) matrix \cite{CKM},
and, conversely, to discover new physics.
What one could call ``the era of precision CKM experiments''
has been started by the measurements of
the CP violating asymmetry in $B_d \rightarrow \psi K$ made at Babar
\cite{Babar} and Belle \cite{Belle}.
Combining their results one obtains
$\sin{2 \beta} = 0.73 \pm 0.05 (\text{stat})$, 
where $\beta = \arg{\left(- V_{cd} V_{cb}^\ast/ V_{td} V_{tb}^\ast\right)}$
(both experiments include, in addition,
systematic errors of around $0.035$).

Many other tests will be enabled by the experiments
currently taking place at Babar and Belle.
Among them,
there will be an interesting class of experiments probing
the unusual combination $\sin{(2 \beta + \gamma)}$,
where $\gamma = \arg{\left(- V_{ud} V_{ub}^\ast/ V_{cd} V_{cb}^\ast\right)}$.
One possibility concerns decays based on the quark-level decay
$b \rightarrow u c s$.
The first such proposal is contained in an article by
Gronau and London (GL),
and it requires the measurement of the time dependent decays 
$B_d(t) \rightarrow D^0 K_S$ and 
$B_d(t) \rightarrow \overline{D^0} K_S$ \cite{GL}.
Because this idea was combined with the extraction of $\gamma$ from
the rates of $B_d \rightarrow
\{ D^0, \overline{D^0}, D_{cp} \} K_S$,
their point is sometimes overlooked (see, however, ref.~\cite{BLS}).
Nevertheless,
these decays involve the difficult task of identifying the
flavor of the $D^0$ or $\overline{D^0}$ mesons in the final state,
which, moreover, may mix with each other \cite{BtoD}.
This has prompted Kayser and London (KL) to extend the idea
to the decays $B_d(t) \rightarrow D^{\ast \ast 0} K_S$ \cite{KL,AS}.
Another possibility concerns decays based on the quark-level decay
$b \rightarrow u c d$.
Dunietz pointed out that $\sin(2 \beta + \gamma)$ can
also be determined from the time dependent decays
$B_d(t) \rightarrow D^{\ast \pm} \pi^\mp$ \cite{Dun98},
while London, Sinha and Sinha (LSS) stressed that the decays
$B_d(t) \rightarrow D^{(\ast)\pm} \left\{ \rho^\mp, a_1^\pm \right\}$
may have some advantages,
despite the fact that an angular analysis becomes necessary \cite{LSS}.
The nice feature of all these decays is that they involve only
tree level diagrams and, thus, are not subject to penguin pollution.

Given that precise measurements of $\sin{2 \beta}$ and
$\sin(2 \beta + \gamma)$ will become available in the next few years,
it is important to ask what one will learn from them.
This is the question we address here.
Our analysis differs from previous ones in that we are not proposing
any method in particular.
Rather,
we are interested in what one can learn about the fundamental
physics,
once any (or several) such method(s) is (are) implemented.
We will focus on the following three questions:
\begin{itemize}
\item Under which conditions will we be able to learn something about
the weak phase $\gamma$, if it lies within its current SM allowed value?
\item How does a measurement of $\sin{(2 \beta + \gamma)}$
help us to constrain new physics?
\item How do strong phases impact the previous questions?
\end{itemize}
We obtain qualitative answers to these questions by looking at a number
of examples, but we do not try to simulate statistical analysis of
prospective data, since that will depend on the precise decay and
method used.

In section~\ref{sec:SM} we assume the SM and analyze the accuracy with which
$\gamma$ can be determined in prospective scenarios.
In subsection~\ref{subsec:current} we define our notation and review
the current status of the SM.
In subsection~\ref{subsec:future} we address the impact that a measurement
of $\sin{(2 \beta + \gamma)}$ is likely to have on our knowledge of
$\gamma$,
if this phase happens to lie within its current SM allowed values.
We will argue that,
if the (one sigma) errors on
$\sin{2 \beta}$ and $\sin{(2 \beta + \gamma)}$
are $0.025$ and $0.1$, respectively,
we are likely to learn very little.
On the other hand,
even a twofold improvement in those errors may
allow us to make an improvement over the present constraints on
$\gamma$.
In those sections,
we ignore the problems brought about by the strong phases.
These are dealt with in subsection~\ref{subsec:strong}.

Section~\ref{sec:new} is devoted to a study of the
constraints imposed on new physics by a measurement of
$\sin{(2 \beta + \gamma)}$.
We stress the importance of the \textit{complex} matrix elements
$V_{td}$ and $V_{ub}$ which, in the usual phase convention,
have phases $\beta$ and $\gamma$, respectively.
The determination of \textit{either} of these,
together with our present knowledge of the other matrix elements,
completes the determination of the CKM matrix.
Here we emphasize the fact that the experimental determination
of the (complex) matrix element $V_{td}$ depends entirely
on $B - \overline{B}$ mixing, which,
because it occurs in the SM through a box diagram,
can be subject to large new physics effects.
We denote this ``mixing'' information by $\tilde{V}_{td}$:
its phase is determined from the CP-violating asymmetry already
measured at Belle and Babar, $\sin 2 \tilde{\beta}$,
and its magnitude,
$|\tilde{V}_{td}|$,
will be well determined once the $B_s - \overline{B_s}$ mass difference
$\Delta m_s$ is measured at Fermilab.
Indeed,
although the extraction of $|\tilde{V}_{td}|$ 
from $\Delta m_d$ is plagued
by large hadronic uncertainties,
it is believed that those uncertainties are under much better control
for the ratio $\Delta m_d/\Delta m_s$ \cite{Kron02}.
Combining these two measurements gives the pair we label
$(\tilde{\rho}, \tilde{\eta})$.
In contrast,
the determination of $V_{ub}$ hinges on decay,
not mixing.
The magnitude of $V_{ub}$ is obtained from semileptonic decays,
but the precision is poor because of hadronic uncertainties.
Here we focus on the determination of $\gamma$ from
the comparison between $\sin{(2 \tilde{\beta} + \gamma)}$
and $\sin 2 \tilde{\beta}$.
Combining these two measurements gives the pair we label
$(\rho, \eta)$.
In this section~\ref{sec:new},
we explore the constraints one can place on a new physics
contribution to $B_d -\overline{B_d}$ mixing
by contrasting the information gathered from mixing,
$\tilde{V}_{td}$,
with that obtained from decay,
$V_{ub}$.
(Note that in section~\ref{sec:SM},
where we assume the SM,
we make no distinction between $\beta$ and $\tilde{\beta}$.
This distinction becomes necessary in the presence of new physics.)

We draw our conclusions in section~\ref{sec:conclusions}.
One of our main points is the following:
in the short term,
a measurement of $\sin{(2 \beta + \gamma)}$ is likely to be more
useful as a probe for new physics than it will be for a better
determination of the phase $\gamma$ in the SM.

\section{\label{sec:SM}Determining $\gamma$ in the SM with
$\sin{(2 \beta + \gamma)}$}

\subsection{\label{subsec:current}Current status}

For convenience,
we introduce the following notation:
\begin{eqnarray}
\phi &=& 2 \beta+\gamma,
\nonumber\\
a_\beta &=& \sin{2 \beta},
\nonumber\\
a_\phi &=& \sin(2 \beta + \gamma).
\end{eqnarray}
By combining the experimental bounds on
$|V_{ub}/V_{cb}|$,
$|\epsilon_K|$,
$x_d$,
and $x_s$,
Ali and London obtained in early 2000 the following
95\% C.L. ranges for the Standard Model \cite{Ali00},
\begin{eqnarray}
16^\circ \leq \beta \leq 34^\circ
\ \ & \Rightarrow &\ \ 
0.53 \leq \sin{2 \beta} \leq 0.93,
\nonumber\\
38^\circ \leq \gamma \leq 81^\circ
\ \ & \Rightarrow &\ \ 
0.38 \leq \sin^2{\gamma} \leq 0.98\ ,
\label{bg-limit-2}
\end{eqnarray}
leading to
\begin{equation}
70^\circ \leq \phi \leq 149^\circ
\ \ \Rightarrow \ \ 
\left\{
\begin{array}{l}
0.94 \leq \sin{\phi} \leq 1.00,
\ \ \ \mbox{if}\ \phi\  \mbox{is in the 1st quadrant,}\\
0.51 \leq \sin{\phi} \leq 1.00,
\ \ \ \mbox{if}\ \phi\  \mbox{is in the 2nd quadrant.}
\label{phi-limit-2}
\end{array}
\right.
\end{equation}
Perhaps surprisingly,
given the rather different assumptions and statistics used,
a later analysis by H\"{o}cker, Lacker, Laplace and Le Diberder \cite{Hock01},
including the then world average $\sin{2 \beta} = 0.793 \pm 0.102$,
reached very similar 95\% C.L. bounds
\begin{eqnarray}
18^\circ \leq \beta \leq 31^\circ
\ \ & \Rightarrow &\ \ 
0.59 \leq \sin{2 \beta} \leq 0.88,
\nonumber\\
37^\circ \leq \gamma \leq 80^\circ
\ \ & \Rightarrow &\ \ 
0.36 \leq \sin^2{\gamma} \leq 0.97\ ,
\label{bg-limit-3}
\end{eqnarray}
leading to
\begin{equation}
73^\circ \leq \phi \leq 142^\circ
\ \ \Rightarrow \ \ 
\left\{
\begin{array}{l}
0.95 \leq \sin{\phi} \leq 1.00,
\ \ \ \mbox{if}\ \phi\  \mbox{is in the 1st quadrant,}\\
0.61 \leq \sin{\phi} \leq 1.00,
\ \ \ \mbox{if}\ \phi\  \mbox{is in the 2nd quadrant.}
\label{phi-limit-3}
\end{array}
\right.
\end{equation}
Since we want to illustrate what can be learned from measurements
of $a_\phi$ and $a_\beta$,
we are not too concerned about the precise values of these bounds which,
moreover, will have been improved by the time the analysis is
performed.
(For example, the 95\% C.L. ranges for $\sin{2 \beta}$ quoted
on the first line of Eq.~\ref{bg-limit-3} are almost
identical to the latest experimental limits from
Babar \cite{Babar} and Belle \cite{Belle}.)
Notice that
both the current lower bound on $x_s$ and a lower bound on
$\sin{(2 \beta + \gamma)}$ chip away at the values of
$\gamma$ around $90^\circ$.

In the Standard Model,
$2 \beta$ is known to be in the first quadrant.
However, $2 \beta + \gamma$ can be in the first or in the second
quadrant.
Therefore we must consider 2 cases in the extraction of $\gamma$:
\begin{itemize}
\item Case A: the measurement of $a_\phi$ excludes
values larger than $0.95$ at 95\% C.L.;
\item Case B: the measurement of $a_\phi$ does {\em not} exclude
values larger than $0.95$ at 95\% C.L.
\end{itemize}
In the first case,
$\phi$ is in the second quadrant and the value of $\gamma$ is obtained by
\begin{equation}
\gamma_2 = \pi - \arcsin{(a_\phi)} - \arcsin{(a_\beta)}.
\label{gamma2}
\end{equation}
In the second case,
either $\phi$ is in the second quadrant,
with $\gamma$ given by Eq.~(\ref{gamma2}),
or, alternatively,
$\phi$ is in the first quadrant,
with $\gamma$ given by
\begin{equation}
\gamma_1 = \arcsin{(a_\phi)} - \arcsin{(a_\beta)}.
\label{gamma1}
\end{equation}

We are also interested in the observable
\begin{equation}
\epsilon^\prime_{\text{B2}} = a_{\phi} - a_{\beta}=
\sin{(2 \beta + \gamma)} - \sin{2 \beta},
\label{DCP}
\end{equation}
which measures direct CP violation \cite{Wolf01}.
This is part of a general class of observables of the type
$\eta_f \lambda_f - \eta_g \lambda_g$ \cite{pages} where
$f$ and $g$ are CP eigenstates with CP-parities
$\eta_f$ and $\eta_g$,
and,
as usual,
\begin{equation}
\lambda_{f(g)} = \frac{q}{p}\, 
\frac{A[\overline{B^0} \rightarrow f(g)]}{A[B^0 \rightarrow f(g)]}.
\end{equation}
This is a very interesting class of observables because
it exhibits direct CP violation {\em without} the need for
strong phases.
Another observable which belongs to this class is
the kaon parameter $\epsilon^\prime_K$ \cite{Wolf01,pages}.

\subsection{\label{subsec:future}Examples of the impact of future experiments}

Based on recent simulation studies \cite{phi-error},
we assume that the errors on  $\sin{(2 \beta+\gamma)}$ and 
$\sin{2 \beta}$ satisfy the relation
$\sigma_{a_\phi} = 4 \sigma_{a_\beta}$.
Given the errors in Refs.~\cite{Babar} and \cite{Belle},
we expect that errors of order 
$\sigma_{a_\beta} = 0.025$ and $\sigma_{a_\phi} = 0.1$
(corresponding to an integrated luminosity of
around $700 \text{fb}^{-1}$ or $800 \text{fb}^{-1}$)
might be achieved in 2004,
combining the integrated luminosities of the two B factories.
We will show that,
assuming such errors,
the upcoming measurement of $\sin{(2 \beta + \gamma)}$
might be much more effective at uncovering large new
physics effects, than it will be in constraining $\gamma$
(if the value of this phase happens to lie in its SM allowed range).
We will also show that
interesting constraints on a SM value for $\gamma$ are possible if
the errors are reduced by a factor of two,
to $\sigma_{a_\beta} = 0.0125$,
for $a_\beta$,
and $\sigma_{a_\phi} = 0.05$, for $a_\phi$.

Here and throughout the rest of the article,
all ranges quoted are $2 \sigma$ corresponding approximately to 95\% C.L.,
and we will not concern ourselves with experimental ranges for
$a_\phi$ and $a_\beta$ outside the interval (-1,1).

We will follow the following strategy:
\begin{itemize}
\item assume some central value for $a_\beta$, taken from its currently
allowed range.
We will assume that ${a_\beta}_{cv}=0.75$
(we will only mention in passing the possibilities that
${a_\beta}_{cv}=0.65$ and ${a_\beta}_{cv}=0.85$);
\item assign an error of $\sigma_{a_\beta} = 0.025$ to this measurement (later
we will also use $0.0125$);
\item assume that the true value for $\beta$ is given by
$2 \beta_{\text{true}} = \arcsin{{a_\beta}_{cv}}$;
\item assume some true value for $\gamma$ ($\gamma_{\text{true}}$).
Combining this with $\beta_{\text{true}}$, we can calculate
the ``true'' value for $\sin{(2 \beta + \gamma)}$. We will assume
that this coincides with its experimentally determined central value,
${a_\phi}_{cv}$;
\item assign an error of $\sigma_{a_\phi} = 0.1$ to this measurement (later
we will also use $0.05$);
\item use the ``experimental'' 95\% C.L. ranges of
${a_\beta}_{cv} \pm 2 \sigma_{a_\beta}$ and
${a_\phi}_{cv} \pm 2 \sigma_{a_\phi}$,
together with Eqs.~(\ref{gamma2}) and (\ref{gamma1}),
to determine the values of $\gamma_{{\text{output}}}$
extracted from ``experiment''.
\end{itemize}

\begin{figure}[htb]
\includegraphics*[height=7cm]{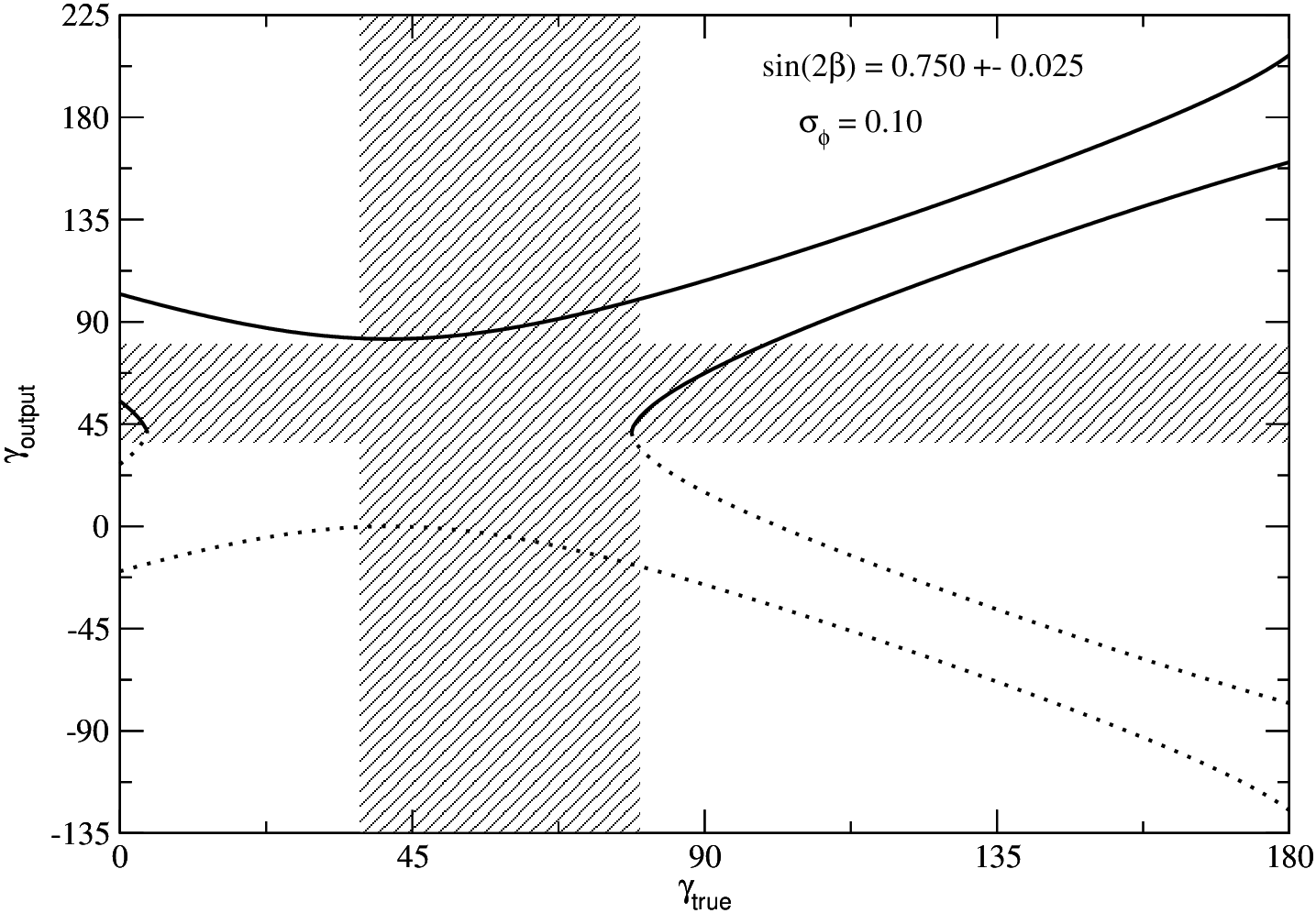}
\caption{\label{figure1}The ``experimentally'' determined
(output) values for $\gamma$,
as a function of its ``true'' value.
The curves correspond to a $2 \sigma$ deviation in
\textit{both} $\sin{2 \beta}$ and $\sin{(2 \beta + \gamma)}$.
The allowed region between the dotted (filled) curves corresponds
to $2\beta+\gamma$ in the first quadrant (second) quadrant.
The hatched horizontal and vertical regions correspond to the
currently allowed range,
$37^\circ \leq \gamma \leq 80^\circ$.
Here we assume that $\sin{2 \beta} = 0.750 \pm 0.025$ and that
the error on $\sin{(2 \beta + \gamma)}$ is $0.1$.
}
\end{figure}

The results are shown in FIG.~\ref{figure1}.
Of course,
the solution $\gamma_{\text{output}} = \gamma_{\text{true}}$
is included in the figure.
Less obvious is the inclusion of the solution
$\gamma_{\text{output}} =
\pi - 2 \arcsin{{a_\beta}_{cv}} - \gamma_{\text{true}}$,
which has to do with the discrete ambiguities to be discussed
in subsection~\ref{subsec:strong}.
The point is that $\sin{\phi} = \sin{(\pi - \phi)}$ and,
therefore,
a measurement of $a_\phi$ is invariant under the transformation
$\gamma \rightarrow \pi - 4 \beta - \gamma$.

Under the assumptions described,
FIG.~\ref{figure1} tells us that,
if $\gamma$ lies within its SM allowed range,
a measurement of $\sin{(2 \beta+\gamma)}$ will essentially not
help us in constraining $\gamma$ any further.
Of course,
there is no good reason to take the central value of
$a_\beta$ to determine the true value of $\beta$;
nor is there any good reason to assume that
the central value determined experimentally for
$a_\phi$ will turn out to correspond to the true
values for $\beta$ and $\gamma$.
In fact, the measurement may hit the tail of the statistics.
Also,
the limiting curves in FIG.~\ref{figure1} correspond to rather
conservative bounds,
since they were found using the extreme values
on both $a_\beta$ and $a_\phi$.
This does not, however,
affect the qualitative conclusions we will draw.
Changing the central value ${a_\beta}_{cv}$ to $0.65$
(0.85) would alter FIG.~\ref{figure1} only slightly
and would would not change our conclusion that,
if the SM holds,
essentially no new information will be gained.

This is one of our main points:
with the errors of $\sigma_{a_\beta} = 0.025$ and
$\sigma_{a_\phi} = 0.1$,
we are unlikely to gain new information on $\gamma$.
Nevertheless,
as we will discuss in section~\ref{sec:new},
such a measurement is useful in constraining new physics.

\begin{figure}[htb]
\includegraphics*[height=7cm]{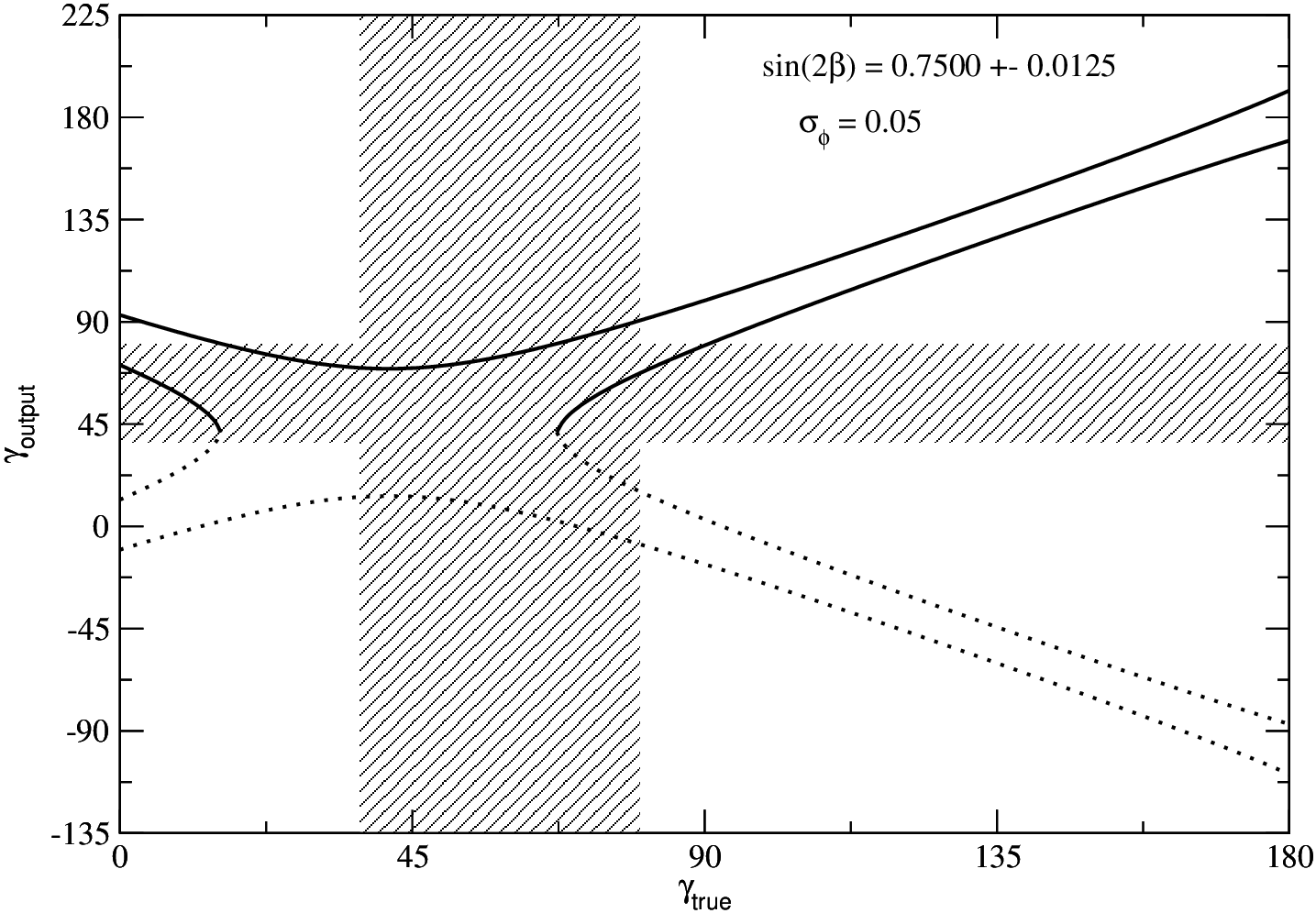}
\caption{\label{figure2}Same as FIG.~\ref{figure1},
but with $\sin{2 \beta} = 0.7500 \pm 0.0125$ and with
an error on $\sin{(2 \beta + \gamma)}$ of $0.05$.}
\end{figure}

The situation improves dramatically as the experimental errors
get smaller.
We illustrate this point in FIG.~\ref{figure2},
were we take ${a_\beta}_{cv} = 0.75$ and we
consider a factor of two improvement in the errors:
$\sigma_{a_\beta} = 0.0125$, $\sigma_{a_\phi} = 0.05$,
requiring an upgrade of the B factory.

Let us illustrate the various possibilities with a few examples.
In Example~1 we assume that the measurements yield
the 95\% C.L. ranges of
$0.7 \leq a_\beta \leq 0.8$ and
$0.6 \leq a_\phi \leq 1.0$ (this corresponds to  FIG.~\ref{figure1} with
$\gamma_{\text{true}} \sim 78^\circ$).
The possibility that $\phi$ is in the first quadrant leads to
a lower limit $\gamma \geq -16^\circ$,
while the second quadrant leads to an upper limit
$\gamma \leq 99^\circ$.
Moreover,
the direct CP-violating parameter $\epsilon^\prime_{\text{B2}}$ is consistent
with zero.
If the experimental results turn out to be as in this example,
we learn absolutely nothing within the SM.
(However,
as we will see in section~\ref{sec:new},
the upper limit could constrain some extreme new physics.)

In Example~2 we assume that the prospective 95\% C.L. experimental
ranges are
$0.725 \leq a_\beta \leq 0.775$ and
$0.7 \leq a_\phi \leq 0.9$ (this corresponds to FIG.~\ref{figure2} with
$\gamma_{\text{true}} \sim 78^\circ$).
For $\phi$ in the first quadrant,
Eq.~(\ref{gamma1}) gives $-6^\circ \leq \gamma \leq 18^\circ$;
for $\phi$ in the second quadrant,
Eq.~(\ref{gamma2}) gives $65^\circ \leq \gamma \leq 89^\circ$.
When compared with the currently allowed range for $\gamma$
in Eq.~(\ref{bg-limit-3}),
we are able to exclude the region $37^\circ \leq \gamma \leq 65^\circ$.
Clearly this comes about because of the upper bound on $a_\phi$;
as this upper bound comes away from $1.0$,
we exclude more and more of the low values of $\gamma$.
While we learn about $\gamma$,
$\epsilon^\prime_{\text{B2}}$ remains consistent with zero.

The opposite could also occur.
To illustrate this point,
let us consider Example~3 where we assume the measurements to yield
the 95\% C.L. ranges
$0.725 \leq a_\beta \leq 0.775$
and $0.8 \leq a_\phi \leq 1.0$
(this corresponds to FIG.~\ref{figure2} with 
$\gamma_{\text{true}} \sim 67^\circ$).
In this case we obtain $2^\circ \leq \gamma$
for $\phi$ in the first quadrant,
and $\gamma \leq 80^\circ$
for $\phi$ in the second quadrant.
Here,
while we gain no information on $\gamma$,
we will have a signal of direct CP violation
because $0.025 \leq \epsilon^\prime_{\text{B2}} \leq 0.275$.

It is also possible that, in the presence of new physics,
$\gamma_{\text{true}}$ is not consistent with
the constraints from  Eq.~(\ref{bg-limit-3}).
As Example~4 let us consider $\gamma_{\text{true}}=105^\circ$
in FIG.~\ref{figure1}.
This corresponds to the prospective 95\% C.L. ranges
$0.7 \leq a_\beta \leq 0.8$
and $0.25 \leq a_\phi \leq 0.65$.
These lead to
$-39^\circ \leq \gamma \leq -4^\circ$,
for $\phi$ in the first quadrant,
and to $86^\circ \leq \gamma \leq 121^\circ$,
for $\phi$ in the second quadrant.
This would be an indication of physics beyond the SM,
although,
as discussed below in subsection~\ref{subsec:strong},
the ambiguity induced by the strong phase might prevent a definitive
conclusion.
To phrase our conclusion differently:
the measurement of $\sin{(2 \beta + \gamma)}$ could,
in principle,
distinguish values of $\gamma$ consistent with
the Standard Model from values of $\gamma$ requiring new physics.

\subsection{\label{subsec:strong}The impact of strong phases}

Thus far we have assumed that a clean measurement of
$\sin{(2 \beta + \gamma)}$ will be available.
However,
the presence of strong phases introduces discrete ambiguities
which we will now discuss.
The Dunietz \cite{Dun98} and KL \cite{KL} methods involve final
states which, although they are not CP eigenstates,
can be accessed by both $B^0$ and $\overline{B^0}$.
Moreover,
as pointed out above,
these decays involve only one weak phase because they are
driven by the purely tree-level quark decay schemes
$b \rightarrow u c s$ and $b \rightarrow u c d$,
respectively.
The importance of decays with these characteristics was first pointed
out by Aleksan, Dunietz, Kayser, and Le Diberder \cite{ADKL},
who showed that measuring the four decays 
$\{B^0, \overline{B^0}\} \rightarrow \{f, \bar f\}$
enables the determination of
\begin{equation}
s_\pm \equiv  \sin{(\phi \pm \Delta)},
\end{equation}
where $\Delta$ is a strong phase,
and $\phi$ is a weak phase (which coincides with $2 \beta + \gamma$
in the Dunietz \cite{Dun98} and KL \cite{KL} methods).
Unfortunately,
knowing $s_\pm$ does not in general determine the sign of
$\cos{(\phi \pm \Delta)}$,
meaning that
\begin{equation}
\sin^2{\phi} = \frac{1}{2} \left[ 1 + s_+ s_- \mp
\sqrt{(1-s_+^2) (1-s_-^2)}
\right],
\label{eq:inversion}
\end{equation}
can be confused with
\begin{equation}
\cos^2{\Delta} = \frac{1}{2} \left[ 1 + s_+ s_- \pm
\sqrt{(1-s_+^2) (1-s_-^2)}
\right].
\label{eq:confusion}
\end{equation}
Thus, we have in general an eightfold ambiguity due to the three
symmetries \cite{BLS}
\begin{subequations}
\begin{eqnarray}
\sin^2{\phi} &\longleftrightarrow& \cos^2{\Delta},
\label{trouble-a1}
\\
\sin{\phi} &\longleftrightarrow& - \sin{\phi},
\\
\sin{\phi} &\longleftrightarrow& \sin{(\pi - \phi)}.
\end{eqnarray}
\end{subequations}
Alternatively,
we may view the eightfold ambiguity as resulting from the operations
\cite{Soffer}
\begin{subequations}
\begin{eqnarray}
\phi \rightarrow \Delta + \pi/2 
& \ \ \mbox{and}\ \ &
\Delta \rightarrow \phi - \pi/2,
\label{trouble-a2}
\\
\phi \rightarrow \phi + \pi,
& \ \ \mbox{and}\ \ &
\Delta \rightarrow \Delta + \pi,
\\
\phi \rightarrow \pi - \phi, 
& \ \ \mbox{and}\ \ &
\Delta \rightarrow -\Delta.
\end{eqnarray}
\end{subequations}
The first of these ambiguities is the most serious if we
cannot constrain the value of $\Delta$.

In fact, the value of $\Delta$ depends on the details of the final
state scattering matrix and, thus, is difficult to predict.
Beneke, Buchalla, Neubert, and Sachrajda \cite{BBNS},
developing on the color transparency arguments of Bjorken \cite{BJ},
have argued that the strong phases in $B_d \rightarrow D$ decays
are small.
On the other hand,
a recent analysis by CLEO of the $B_d \rightarrow D \pi$ decays
\cite{CLEO02} suggests the presence of final state interactions,
with a strong phase $16.5^\circ < \delta_I < 38.1^\circ$
at the $90\%$ C.L.
This result can be understood as a large rescattering
contribution and thus a sizeable strong phase for the color-suppressed
decay to $D^0 \pi^0$,
which then shows up in the isospin analysis as a phase difference
$\delta_I$ between the $I=1/2$ and the $I=3/2$ final states.
As a result the final state $D^+ \pi^-$
is expected to have a non-zero strong phase,
probably smaller than $\delta_I$.

Dunietz's proposal to determine $\sin(2 \beta + \gamma)$ involves
the relative strong phase between the $b \rightarrow c \bar{u} d$
and $\bar b \rightarrow \bar u c \bar d$ contributions
to $B_d (\overline{B_d}) \rightarrow D^{(*)+} \pi^-$.
While these are not color-suppressed,
there may still be significant rescattering from excited final states,
such as $D^{(*)} \rho$.
However,
since the two decay amplitudes are related (essentially
by the interchange of c and u) the same final states are involved
for both of them and, so, the \textit{relative} strong phase
might be expected to be smaller than either one.

Let us now assume that the true value of $\Delta$ is very small but
that we are not allowed to assume this fact in the analysis of the
experiment.
(Indeed,
one does not wish for the extraction of $\sin{(2 \beta + \gamma)}$
to be hampered by theoretical arguments,
especially if it turns out to uncover a potential signal of new
physics.)
Then we have a confusion between $\sin^2{\phi}$
and $\cos^2{\Delta} \sim 1$.
As an example,
consider the possibility that $\phi = 145^\circ$,
corresponding to $\sin{\phi} = 0.57$.
The ambiguity mentioned implies that,
even within the Standard Model,
the experimental results do not allow us to
determine whether $\phi=145^\circ$ and $\Delta = 0$,
or $\phi=90^\circ$ and $\Delta = \pm 55^\circ$.
This is an important problem because,
as illustrated above,
an upper bound on $a_\phi=\sin{(2 \beta + \gamma)}$
can be used to exclude low values of $\gamma$.

\begin{figure}[htb]
\includegraphics*[height=7cm]{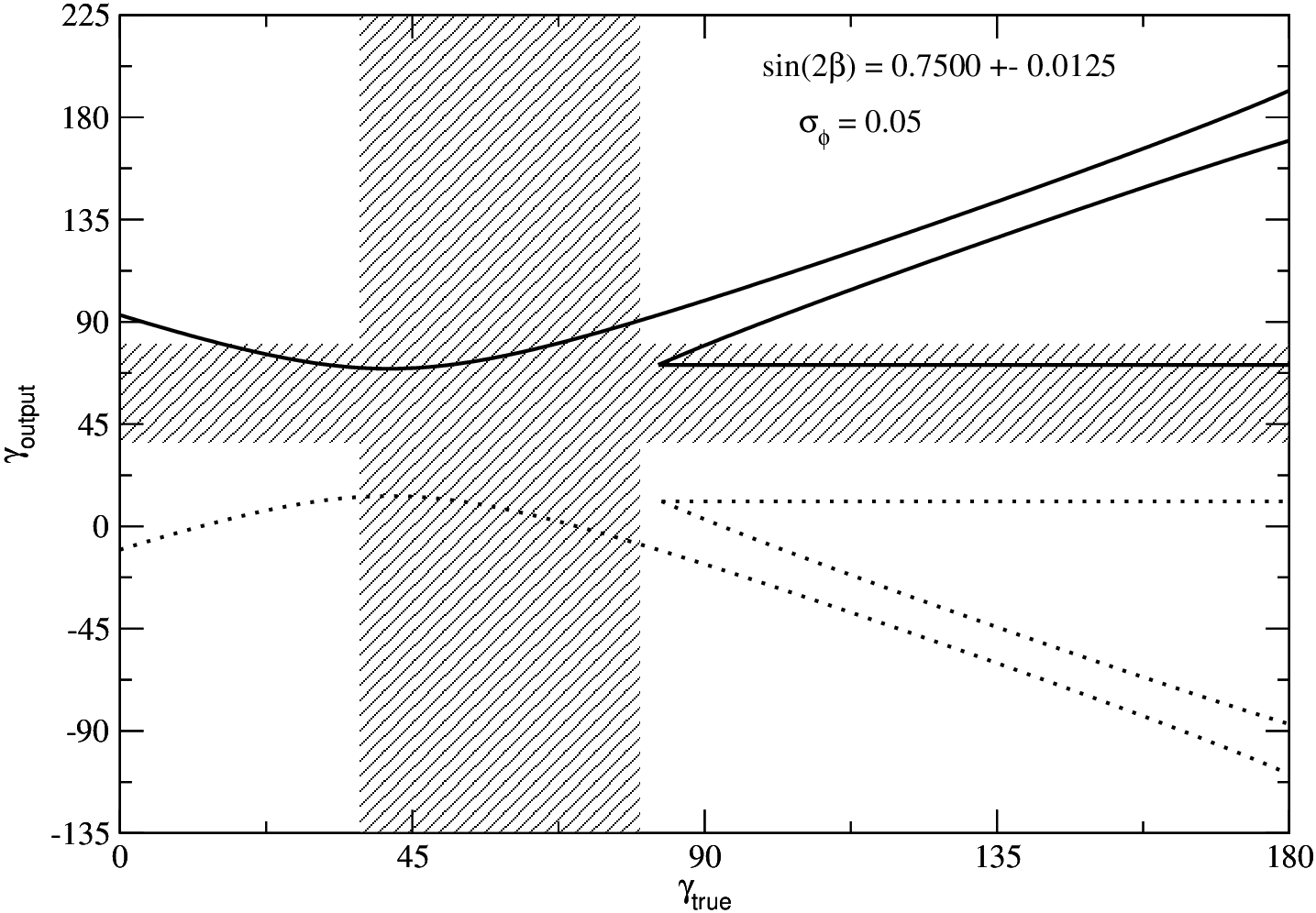}
\caption{\label{figure3}Same as FIG.~\ref{figure2},
but including the ambiguity $\sin^2{\phi} \leftrightarrow \cos^2{\Delta}$,
with $0.9 \leq \cos{\Delta} \leq 1$. 
}
\end{figure}

This ambiguity has no effect on the SM region of FIG.~\ref{figure1}
since there $\sin^2{\phi}$ is consistent with $1$ anyway.
The opposite occurs with FIG.~\ref{figure2},
where it appears that,
for large values of $\gamma_{\text{true}}$,
the analysis ($\gamma_{{\text{output}}}$) can rule out small
values of $\gamma$.
This occurs because there is an upper limit on $\sin^2{\phi}$.
Unfortunately,
the presence of the ambiguity in Eq.~(\ref{trouble-a1})
eliminates this upper limit and the smaller values of $\gamma$ are allowed.
This is illustrated in FIG.~\ref{figure3},
where we have combined the allowed region of FIG.~\ref{figure2} with
the region allowed by the ambiguity
$\sin^2{\phi} \leftrightarrow \cos^2{\Delta}$,
with $0.9 \leq \cos{\Delta} \leq 1$.

As noted in Example~4,
if we allow for the possibility of large new physics,
then $\gamma$ may be outside the SM allowed values.
This can also be seen from FIG.~\ref{figure2} where
the whole SM range is ruled out if $\gamma_{\text{true}} > 91^\circ$.
However,
the exchange in Eq.~(\ref{trouble-a1}) allows
$a_\phi$ to be between $0.9$ and $1$ and,
thus,
most of the SM range is allowed,
as seen in FIG.~\ref{figure3}.
Nevertheless,
since the same interchange leads to a large value for $\Delta$,
which we consider very unlikely,
this could be considered as a strong indication of new
physics.

We should also dispel two common misconceptions.
It is often mentioned that one may remove the ambiguities due
to $\Delta$ by comparing two different final states.
For example, 
we could compare the results in $B_d \rightarrow D^{\pm} \rho^\mp$ 
with those in $B_d \rightarrow D^{\ast \pm} \rho^\mp$,
thus identifying the strong phases.
However,
such a statement carries the hidden assumption that
\textit{the distinction between the two strong phases is 
experimentally feasible}.
Given the expected experimental precision and assuming that the
strong phases are indeed small,
this may not work in practice.

One other common idea concerns the usefulness of a large
final state phase in improving the sensitivity of
a measurement of $\sin{(2 \beta + \gamma)}$ to the phase
$\gamma$.
Recall that,
as we have seen above,
using the current SM ranges for $\beta$ and $\gamma$,
a measurement of $\sin{(2 \beta + \gamma)}$ is not very
sensitive to the different values for $\gamma$.
This is mainly due to the fact that $\sin{(2 \beta + \gamma)}$
lies close to one for a good portion of the allowed range,
where it is less sensitive to small differences in $\gamma$.
One could think that,
given that the measurements involve $\sin{(2 \beta + \gamma \pm \Delta)}$,
a large value for $\Delta$ would overcome this obstacle.
This is not the case.
The point is that,
although the sensitivities of $s_+$ and $s_-$ to the phase $\gamma$
are indeed improved for large values of $\Delta$,
these improvements cancel in the ``inversion procedure'' described in
Eq.~(\ref{eq:inversion}) that leads from these observables to the 
value of $\sin{(2 \beta + \gamma)}$ which we wish to 
know.\footnote{This is easily checked by expanding
$\sin{(2 \beta + \gamma + \delta \gamma \pm \Delta)}$
in powers of $\delta \gamma$ and substituting in
Eq.~(\ref{eq:inversion}).
The result is 
$$\sin^2{(\phi + \delta \gamma)}
\sim
\sin^2{\phi} + \sin{2 \phi}\ \delta \gamma,$$
as it had to be.
Clearly, the term linear in $\delta \gamma$
vanishes whenever $\sin{\phi} \sim 1$ ($\cos{\phi} \sim 0$).}

We conclude that,
in the SM,
the final-state phases cannot improve
the sensitivity of $a_\phi$ to the phase $\gamma$.
Contrary to what one may think,
if these phases are close to zero they are actually an enormous
nuisance,
since in that case,
and whatever the value of $\phi$,
one cannot distinguish the true value of $a_\phi$
from the possibility that $a_\phi \leftrightarrow \cos{\Delta} \sim 1$,
thus precluding the exclusion of small values of $\gamma$.
We have also shown that,
if the final-state phases are large,
the sensitivity of $a_\phi \sim 1$ to the phase $\gamma$ is
not improved at all.
Nevertheless,
in that case,
we may be lucky in disentangling the
strong phases by comparing different channels.

\section{\label{sec:new}Constraining new physics with
$\sin{(2 \beta + \gamma)}$}

Up to now,
we have assumed the SM and concentrated our attention on the capabilities
of a measurement of $\sin{(2 \beta + \gamma)}$ to improve our information
on the CKM phase $\gamma$.
We now turn to the capabilities of this measurement to 
constrain new physics effects.

It is likely that the most precise determination of the
CKM parameters in the next two or three years will come from the
measurements of the CP-violating assymetry in $B_d \rightarrow \psi K$
decays and from the measurements of $\Delta m_d/\Delta m_s$.
These determine the element $\tilde{V}_{td}$ with phase
$\tilde{\beta}$ and magnitude
\begin{equation}
\left| \frac{\tilde{V}_{td}}{V_{ts}} \right|
=
\xi
\sqrt{\frac{\Delta m_d}{\Delta m_s}\, \frac{m_{Bs}}{m_{Bd}}}.
\label{WolfA}
\end{equation}
The factor $\xi$ is 1 in the SU(3) limit and has been calculated
on the lattice to be $1.15 \pm 0.04$ \cite{UKQCD},
but more theoretical work is needed to get the corrections
to the quenched approximation \cite{Kron02}.

\begin{figure}[htb]
\includegraphics*[height=7cm]{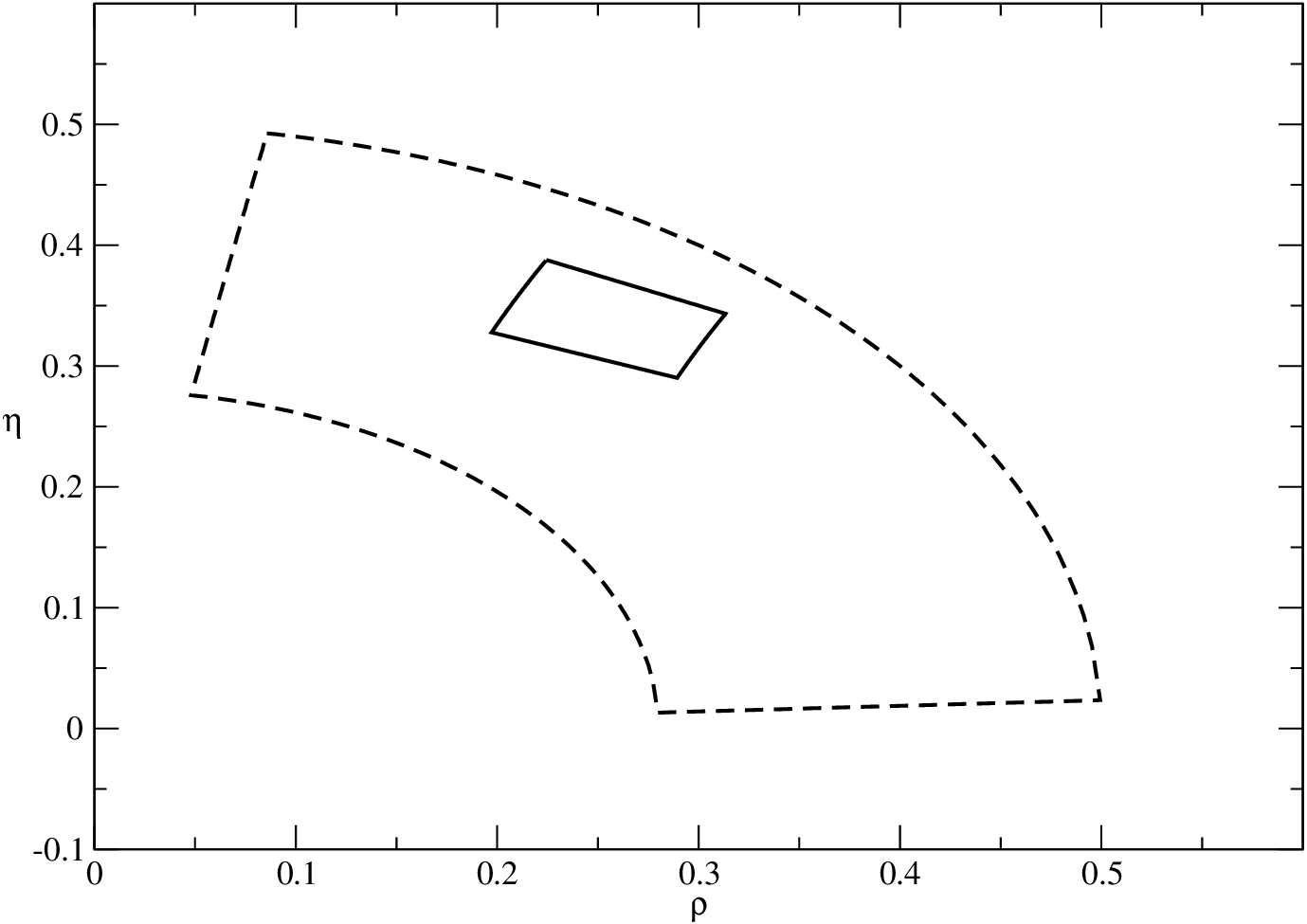}
\caption{\label{figure4}The solid curve shows the allowed region
in the $(\rho, \eta)$ plane based on assumed mixing data with
$\Delta m_s = 20 ps^{-1}$.
The dashed lines show the constraints from assumed decay data.}
\end{figure}

As an example,
we consider the possibility that
$\Delta m_s$ is found to be $20 ps^{-1}$,
$0.7 \leq \sin 2 \tilde{\beta} \leq 0.8$,
and
$1.08 \leq \xi \leq 1.22$,
with ranges corresponding to a 95\% C.L.
We assume that the major error in applying Eq.~(\ref{WolfA}) is the
theoretical error in $\xi$,
since the experimental error in $\Delta m_s$
is expected to be very small,
once it is measured \cite{BPTevatron}.
These two constraints define the region of
$(\tilde{\rho}, \tilde{\eta})$ shown by the
solid line in FIG.~\ref{figure4}.
The case where $\Delta m_s=30 ps^{-1}$ would appear as a similar
region displaced to the right and downward,
but with similar structure.

We now look at the constraints that can be placed on
$(\rho, \eta)$ solely on the basis of $B$ decays,
in contrast to the $\tilde{V}_{td}$-based anaysis,
that depended entirely on mixing.
The magnitude of $V_{ub}$ is determined from semileptonic decays
and its phase $\gamma$ is determined by comparing
$\sin(2 \tilde{\beta}+ \gamma)$ with $\sin 2 \tilde{\beta}$,
as discussed in the previous sections.
For our example,
we assume that $0.28 \leq \sqrt{\rho^2 + \eta^2} \leq 0.5$
at the 95\% C.L.
This is an estimate of the present theoretical error \cite{Hock01},
and we believe that no great progress in reducing this error
is likely to occur in the near future.
To draw the constraints from decay we use
$\sin(2 \tilde{\beta})=0.75$,
${a_\phi}_{cv}=0.98$,
$\sigma_{a_\phi} = 0.1$,
and allow $a_\phi$ to vary within the 95\% C.L. range
$0.78 \leq a_\phi \leq 1$.
This means that,
by construction,
the central value of $\gamma$ corresponds to the central
value of $(\tilde{\rho}, \tilde{\eta})$ in the example above,
while $\tilde{\beta}$ is held fixed at $\arcsin{(0.75)}/2$.
The resulting constraints on $(\rho, \eta)$ from
decay are shown as dashed lines in FIG.~\ref{figure4}
(the limits on $\gamma$ shown in FIG.~\ref{figure4} are slightly
more restrictive than the corresponding ones shown in
FIG.~\ref{figure1} because here we fix the value
of $\tilde{\beta}$).
We note that,
in this case,
the ambiguity from the strong phase disappears,
assuming that it is not too different from zero.

\begin{figure}[htb]
\includegraphics*[height=7cm]{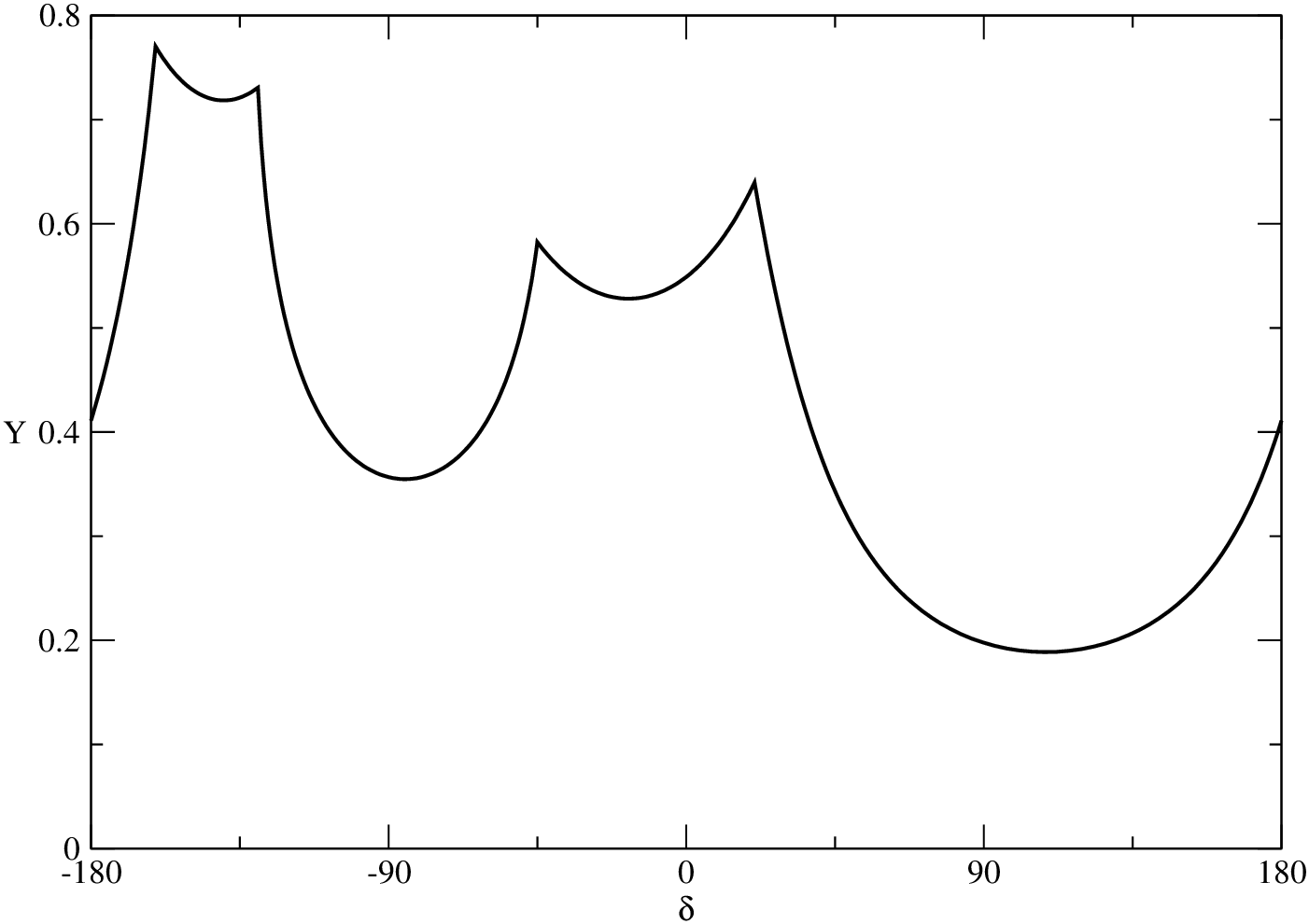}
\caption{\label{figure5}Limits on the new physics contributions
to mixing.
The limits corresponding to $-45^\circ < \delta < 21^\circ$
($-160^\circ < \delta < -129^\circ$)
arises from the upper (lower) bound on $\gamma$.}
\end{figure}

As a result of new physics,
the value from mixing may be different from the value from decay.
As an example,
we consider the possibility that the new physics only shows up 
as a contribution to
the $B_d - \overline{B_d}$ mixing matrix $M_{12}$.
We then write
\begin{equation}
M_{12}
=
M_{12}(\tilde{\rho}, \tilde{\eta})
=
M_{12}(\rho, \eta) - Y e^{i \delta} M_{12},
\end{equation}
where $Y$ measures the magnitude of the new physics as a fraction of
$M_{12}$,
and $\delta$ is the phase of the new physics relative to
$2 \tilde{\beta}$.
Using the central values for $(\tilde{\rho}, \tilde{\eta})$,
each point $(\rho, \eta)$ along the boundary of the allowed region
for decay, shown in FIG.~\ref{figure4},
can be used to calculate values of $Y$ and $\delta$ \cite{Goto96}.
The resulting limits on $Y$ as a function of $\delta$ are
shown in FIG.~\ref{figure5}.
We see that the constraint from $\sin(2 \tilde{\beta} + \gamma)$
provides a constraint on $Y$ for values of $\delta$ in the ranges
$(-160^\circ, -129^\circ)$ and $(-45^\circ, 21^\circ)$.
The cusps in Fig.~\ref{figure5} correspond to the corners of the dashed
curve in Fig.~\ref{figure4} and,
thus,
to $2 \sigma$ deviations in $\xi$ as well as in $a_\phi$.
We have not considered constraints from kaon physics;
for a portion of the region allowed for decays
(that with smaller values of $\eta$),
one would need to assume new physics in $K - \overline{K}$
mixing as well as in $B_d - \overline{B_d}$ mixing.

\begin{figure}[htb]
\includegraphics*[height=7cm]{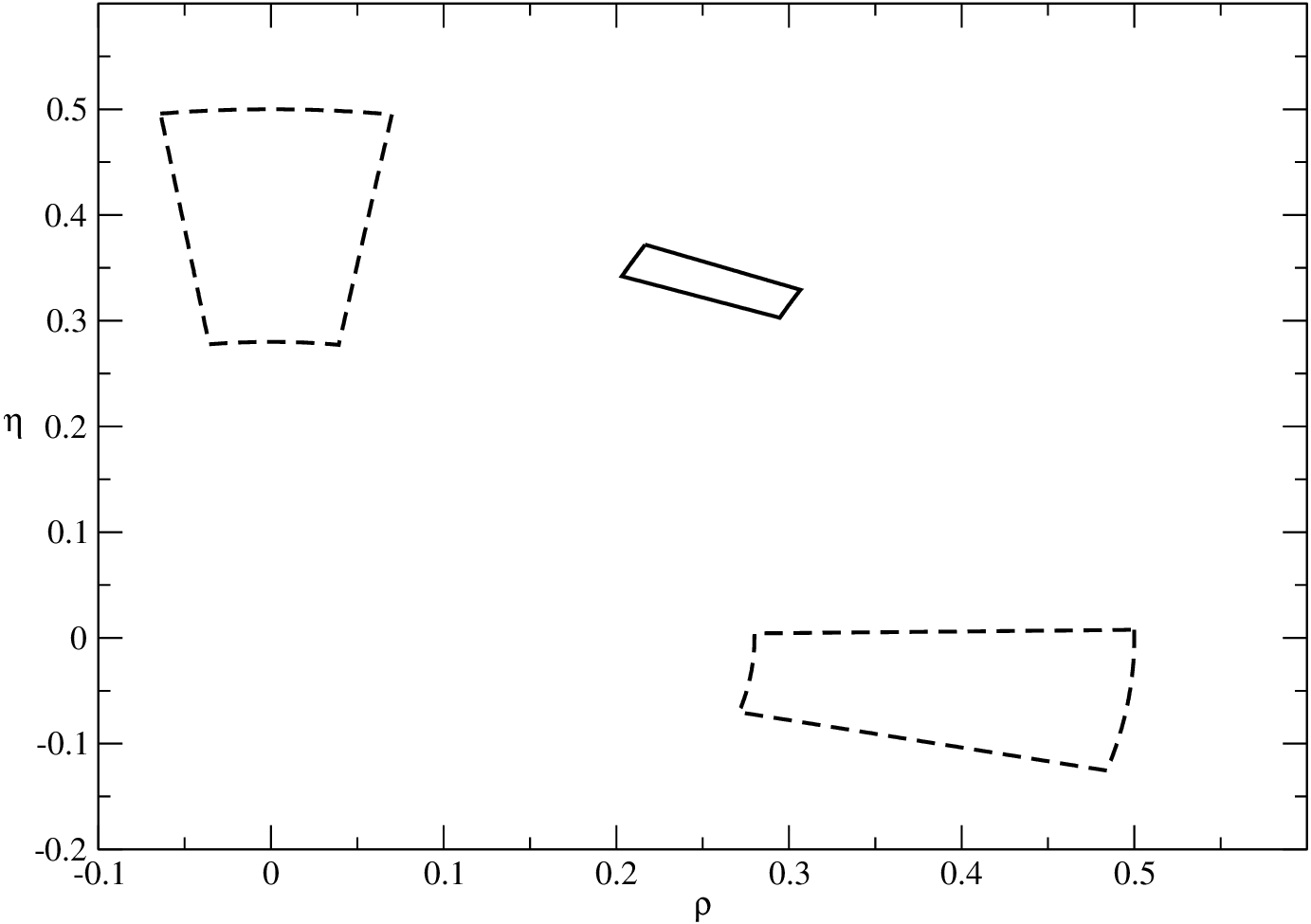}
\caption{\label{figure6}The solid curve shows the allowed region
in the $(\rho, \eta)$ plane based on assumed mixing data with
$\Delta m_s = 20 ps^{-1}$.
The dashed lines show the constraints from assumed decay data.}
\end{figure}

\begin{figure}[htb]
\includegraphics*[height=7cm]{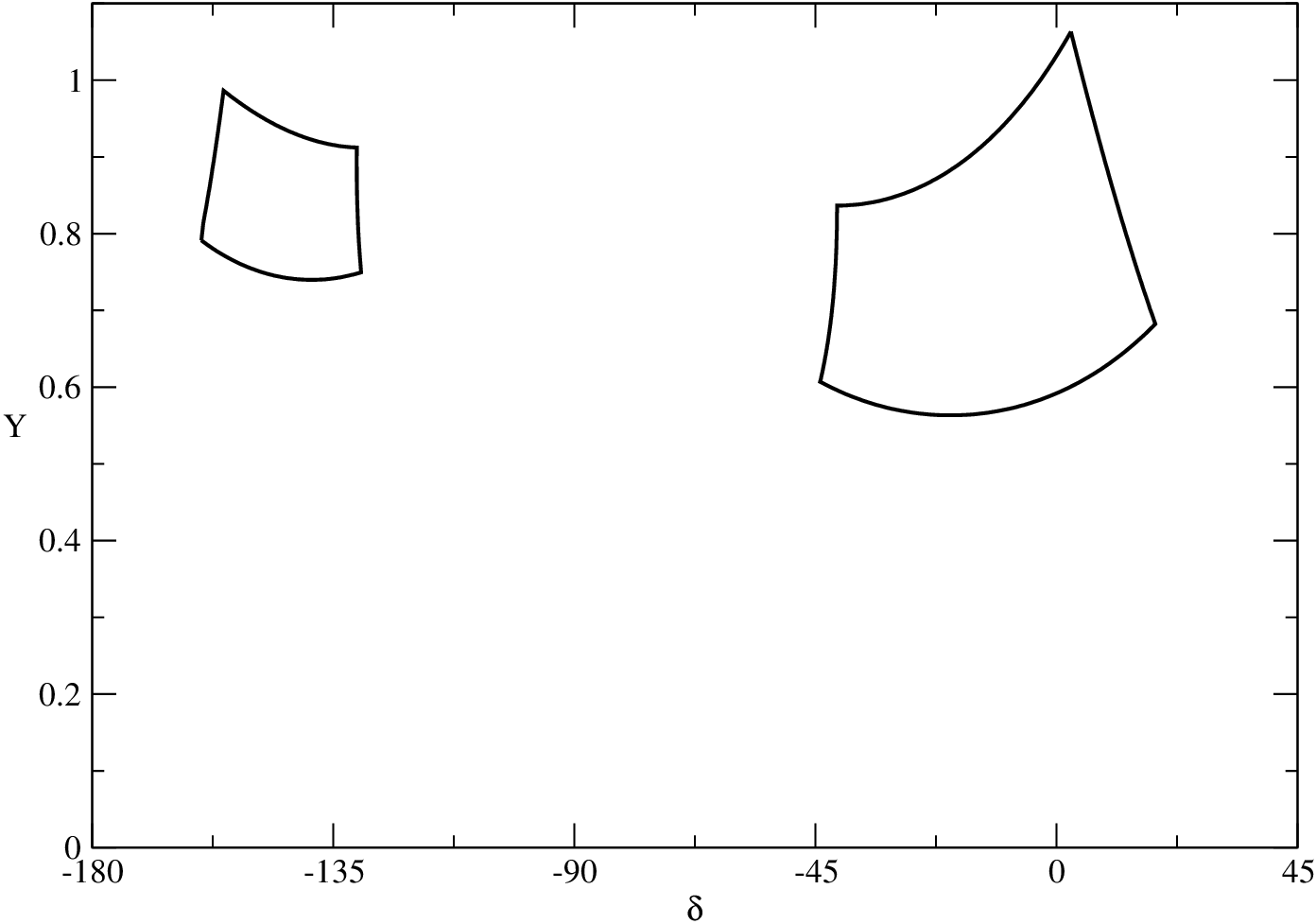}
\caption{\label{figure7}Possible values of the new physics contributions
to mixing.
The bounds corresponding to $-160^\circ < \delta < -130^\circ$
($-44^\circ < \delta < 18^\circ$) arise from $\phi$ in
first (second) quadrant.}
\end{figure}

As mentioned before,
barring the ambiguity due to the strong phase,
smaller errors may allow us to detect new physics, even if
$\gamma$ were not too far outside the range currently allowed
in the SM.
We illustrate this point by repeating the procedure described
above,
but with prospective measurements of
$a_\beta = 0.7500 \pm 0.0125$ and $a_\phi = 0.66 \pm 0.05$
(this corresponds to $\gamma_{\text{true}} \sim 90^\circ$
in FIG.~\ref{figure2}, leading to two separate regions
for $\gamma_1$ and $\gamma_2$).
In this case,
we can see from FIG.~\ref{figure6} that the mixing infered
$(\tilde{\rho},\tilde{\eta})$ region
(bounded by the solid lines)
is completely disjoint from
the decay infered $(\rho,\eta)$ regions
(bounded by the dashed lines).
As a result,
$Y$ is different from zero for some value of $\delta$
in the ranges $(-160^\circ, -130^\circ)$ and
$(-44^\circ, 18^\circ)$,
as shown in FIG.~\ref{figure7},
and we would have identified new physics.
However,
we must consider the ambiguity due to
$\sin^2{\phi} \leftrightarrow \cos^2{\Delta}$.
In  this particular case,
that leaves us with the option of believing in a
rather large value for $Y$ or, alternatively,
admitting that there is no new physics but
the strong phase is large,
between $45^\circ$ and $52^\circ$.
We may hope that,
as we learn more about the systematics of these decays,
the arguments against such a large strong phase may become
decisive.

\section{\label{sec:conclusions}Conclusions}

In this paper we have considered what can be learned from
prospective experiments on $\sin{(2\beta+\gamma)}$.
We have not tried to simulate the future error analysis but,
rather,
by considering a few examples,
have reached qualitative conclusions.
Assuming the standard model is correct,
one can find significant constraints on the value of
$\gamma$ only if the error is $\sigma_{a_\phi} = 0.05$ or less and,
even then,
stronger constraints are likely from the
$\sin{2 \beta}$ and $\Delta m_s$ experiments.
On the other hand,
if we allow for the possibility that there
may be new physics contributions to
$B_d - \overline{B_d}$ mixing,
then comparing the results from experiments probing
$\sin{(2 \tilde{\beta}+\gamma)}$ with the results obtained from mixing
(namely, $\sin{2 \tilde{\beta}}$ and $\Delta m_d/\Delta m_s$)
can provide significant constraints on the new physics contribution,
and may even give an indication of the presence of such contributions.

\begin{acknowledgments}
J.\ P.\ S.\ is greatly indebted to the kind hospitality of
SLAC's Theory Group, where a portion of this work was done.
His work is funded by FCT under POCTI/FNU/37449/2001.
The work of A.\ S.\ is funded by the Department of Energy under
Grant No.\ DE-FG 03-93 ER 40788.
The research of L.\ W.\ and F.\ W.\ is supported in part by the
Department of Energy under Grant No.\ DE-FG 02-91 ER 40682.

\end{acknowledgments}


\end{document}